\documentclass[%
 reprint,
superscriptaddress,
 amsmath,amssymb,
 aps,
 prl
]{revtex4-2}

\usepackage{graphicx}
\usepackage{bm}
\usepackage{bbm}
\usepackage{mathrsfs}
\usepackage{color}
\usepackage{hyperref}
\usepackage[english]{babel}      

\begin{document}

\preprint{APS/123-QED}

\title{Balancing Mobility Behaviors to avoid Global epidemics from Local Outbreaks}

\author{Pablo Valga\~n\'on}
 \affiliation{Department of Condensed Matter Physics, University of Zaragoza, 50009 Zaragoza (Spain).}
\affiliation{GOTHAM lab, Institute for Biocomputation and Physics of Complex Systems (BIFI), University of Zaragoza, 50018 Zaragoza (Spain)}

\author{Antonio Brotons}
 \affiliation{Department of Condensed Matter Physics, University of Zaragoza, 50009 Zaragoza (Spain).}
 
\author{David Soriano-Pa\~nos}
 \affiliation{Departament d'Enginyeria Inform\`atica i Matem\`atiques, Universitat Rovira i Virgili, 43007 Tarragona (Spain)}
\affiliation{GOTHAM lab, Institute for Biocomputation and Physics of Complex Systems (BIFI), University of Zaragoza, 50018 Zaragoza (Spain)}

\author{Jesús G\'omez-Garde\~nes}
 \affiliation{Department of Condensed Matter Physics, University of Zaragoza, 50009 Zaragoza (Spain).}
 
\affiliation{GOTHAM lab, Institute for Biocomputation and Physics of Complex Systems (BIFI), University of Zaragoza, 50018 Zaragoza (Spain)}

\begin{abstract}
Human interactions and mobility shape epidemic dynamics by facilitating disease outbreaks and their spatial spread across regions. Traditional models often isolate commuting and random mobility as separate behaviors, focusing either on short, recurrent trips or on random, exploratory movements. Here, we propose a unified formalism that allows a smooth transition between commuting and exploratory behavior based on travel and return probabilities. We derive an analytical expression for the epidemic threshold, revealing a non-monotonic dependence on recurrence rates: while recurrence tends to lower the threshold by increasing agent concentration in high-contact hubs, it counterintuitively raises the invasion threshold in low-mobility scenarios, suggesting that allowing recurrence may foster local outbreaks while suppressing global epidemics.  These results provide a comprehensive understanding of the interplay between human mobility patterns and epidemic spread, with implications for containment strategies in structured populations.

\end{abstract}
\maketitle

{\em Introduction.-} The rapid globalization of the past century, driven by increased urbanization and advancements in fast, efficient transport, has accelerated the global spread of epidemics. In this context, epidemic modeling~\cite{diekmann.mathematical.tools, anderson1991infectious} has become a crucial tool for informing containment strategies during health emergencies~\cite{hufnagel2004forecast, brizuela2021understanding}. While early compartmental models~\cite{kermack1927contribution} offered initial insights into epidemic dynamics, the spatial aspect introduced by metapopulation frameworks~\cite{Mollison+1972+579+614, HOSONO.TRAVELING.WAVES} highlighted human mobility as a key driver in disease transmission~\cite{COSNER2009550}. 

Current epidemic models integrate mobility effects, considering both commuting~\cite{10.1371/journal.pone.0083002, le2022high, jia2020population, gatto2020spread} and global migration~\cite{rvachev1985mathematical, brockmann2013hidden, doi:10.1073/pnas.0510525103}. The reliability of these models depends on the accurate incorporation of human travel patterns, which are influenced by factors such as the length of stay, destination preferences, and the types of data sources utilized. At this point, based on the various time scales involved in the travel and transmission of disease, one usually has to choose between exploratory agents with access to any part of the system or a more conservative pattern described by commuters.  

In cases of migration and long-term travel, an exploratory (Eulerian) framework is generally preferred, capturing movement as a random walk~\cite{PhysRevLett.92.118701, PhysRevE.85.056115, MASUDA20171}, focusing on the location of the agents at each moment. Conversely, for urban systems with recurrent, short trips, a commuting (Lagrangian) approach is more suitable. In this latter case, agents reside in fixed locations but frequently travel to other patches, facilitating disease transmission across regions. This framework allows either simplifying contagion processes by coupling the infection forces between subpopulations~\cite{keeling.spatial.coupling.2002, gomez2018critical, balcan.multiscale.2009} or explicitly replicating the movement of agents~\cite{Soriano-Paños_2020, POLETTO201341}.

Although some studies incorporate both global travel and commuting, they typically assign these behaviors to distinct types of agents. This assumption overlooks that commuters typically display hybrid mobility patterns, as their visitation frequencies follow a Zipf's law~\cite{gonzalez2008understanding}. To bridge this gap, here we develop a formalism that unifies both descriptions of mobility in a unique model that can seamlessly transition from recurrent travel to random walks and is consistent in the way it characterizes agent interactions and disease transmission.

\begin{figure}[t!]
\includegraphics[width=0.79\linewidth]{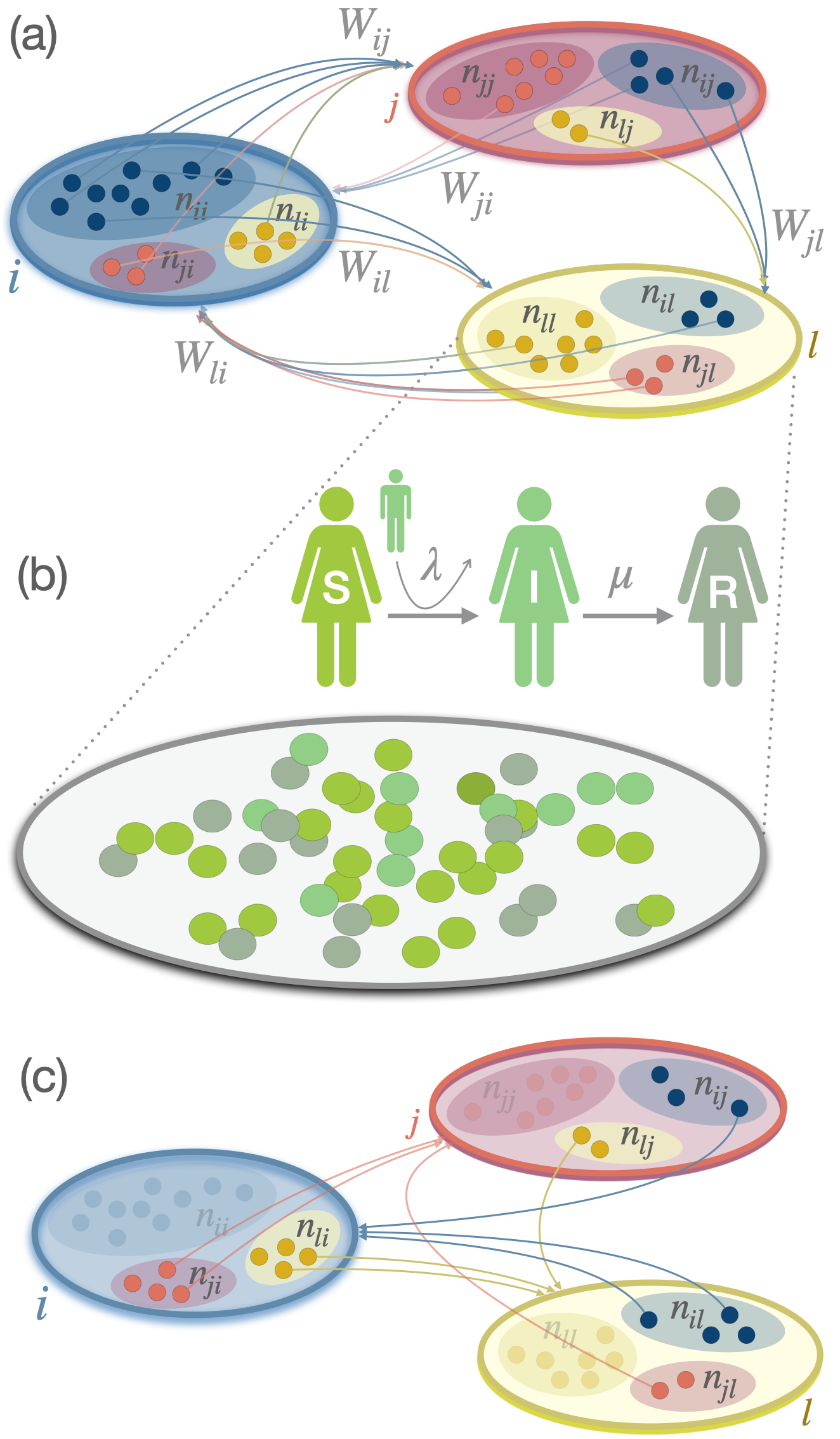}
\caption{\label{Fig1} Schematic description of the three dynamical stages of each time step of the model. Panel (a) shows the Mobility stage in which agents can travel to an adjacent patch (according to matrix ${W}$) with probability $p$. In the Interaction stage (b) the compartmental dynamics (here the SIR model with infection and recovery probabilities $\lambda$ and $\mu$) within each of the patches comes into play. Finally, the Return stage (c) only affects those agents located in a patch other than their home patch, allowing them to return to their residence with probability $r$.}
\end{figure}

{\em Population flows.}- In Figure~\ref{Fig1} we sketch the main features of the reaction-diffusion model here presented. Namely, our model introduces a discrete-time approach,
dividing each time step (day) into three basic stages: movement (M) (Fig.~\ref{Fig1}.a), interaction (I) (Fig.~\ref{Fig1}.b), and return (R) (Fig.~\ref{Fig1}.c). While stage I is responsible for contagions (reactions) occurring inside patches, stages M and R govern the population flows (diffusion) across the metapopulation. In these stages, agents effectively perform a random walk with stochastic resetting ~\cite{10.1093/bioinformatics/bty637,10.1371/journal.pone.0213857,michelitsch2024random}, allowing us to transition between exploratory (random walk) and conservative (commuting) navigations as a function of the relationship between the mobility parameters involved in each stage.

For stage M (Fig.~\ref{Fig1}.a) we define the probability that an agent travels to one of the neighboring patches as $p$. Consequently, $p$ tunes the frequency of movements observed in the metapopulation. In addition, for each of the agents (regardless of their patch of residence) placed at a generic patch $i$, the odds that the destination is patch $j$ is denoted by $W_{ij}$. Conversely, for stage R (Fig.~\ref{Fig1}.c), we introduce a new parameter, $r$, which is defined as the probability that any individual decides to return to their residence rather than staying in their current patch. Thus, by tuning the ratio between $p$ and $r$ our model enables to accommodate different mobility modes. In particular, when $r\ll p$ we replicate exploratory dynamics whereas considering $r\simeq p$ or $r\gg p$ allow us to encode commuting patterns.

Mathematically, we capture population flows by studying the time evolution of $n_{ij}(t)$, defined as the number of agents with residence in patch $i$ staying at patch $j$ at time $t$. Namely, movements during the M stage update the population distribution $\mathbf{n}(t)$ to $\mathbf{n^D}(t)$ through an operator $\mathbbm{D}$ defined as: 
\begin{equation}\label{eq:movement}
    n_{ij}^D(t) = \left(\mathbbm{D} \mathbf{n}(t)\right)_{ij} = (1 - p) n_{ij}(t) + p \sum_l W_{lj}n_{il}(t)\;.
\end{equation}
Analogously, the return operator $\mathbbm{N}$ acts on the latter distribution to redistribute population after the R stage:
\begin{equation}\label{eq:return}
    n_{ij}(t+1) = \left(\mathbbm{N} \mathbf{n}^D(t)\right)_{ij} = \delta_{ij}\sum_l r n^D_{il}(t) + (1 - r) n^D_{ij}(t)\;.
\end{equation}

In a nutshell, the redistribution of agents across patches at each time step can be described as the combined action of the operators $\mathbbm{N}$ and $\mathbbm{D}$ on the set of variables ${n_{ij}(t)}$, i.e. $\mathbf{n} (t+1)=\mathbbm{N}\mathbbm{D}\mathbf{n}(t)$. In fact, by iterating the joint action of $\mathbbm{N}\mathbbm{D}$, one would reach to the stationary distribution of agents across patches $\{n^{\star}_{ij}\}$, corresponding to the leading eigenvector of $\mathbbm{N}\mathbbm{D}$.\\

{\em Epidemic dynamics.-} The former population flows co-exist with the contagion dynamics (stage I), here captured through a compartmental Susceptible-Infected-Recovered (SIR) model. Accordingly, each agent belongs to one of three states ($S$, $I$ and $R$) and can transition between them. In particular, a susceptible becomes infected upon contact with an infected individual with probability $\lambda$ whereas an infected agent has a probability $\mu$ of passing to recovered state per time step.

Contagion and recovery processes take place within patches when, after stage M, agents in different states mix. Let us define $I_{ij}(t)$ and $R_{ij}(t)$ as the number of infected and recovered agents, respectively, who reside in patch $i$ and start time step $t$ at $j$. Then, after stage M, their distribution across patches is given by: $\mathbf{I}^D(t+1) = \mathbbm{D}\mathbf{I}(t)$ and  $\mathbf{R}^D(t+1) = \mathbbm{D}\mathbf{R}(t)$. Thus, by assuming each person contacts all the agents in their current patch, the likelihood that a susceptible agent becomes infected in patch $i$ is:
\begin{equation}
P_i^D (t) = 1 - \prod_{l=1}^{N} \left( 1 - \lambda \frac{I_{li}^D (t)}{n_{li}^D (t)} \right)^{n_{li}^D(t)}\;,
\label{eq:probab}
\end{equation}
so that the number of contagions (new infections) of residents from patch $i$ inside patch $j$ at time $t$ is: 
\begin{equation}
C_{ij}(t) = \left( n_{ij}^D(t) - I_{ij}^D(t) - R_{ij}^D \right) P_j^D(t).
\end{equation}
Finally, the return operator $\mathbbm{N}$ determines the distribution of infected agents after stage R, resulting in:
\begin{equation}
\mathbf{I}(t+1) = (1 - \mu)\mathbbm{N}\mathbf{I}^D(t) + \mathbbm{N}\mathbf{C}(t),
\label{eq:markov1}
\end{equation}
where the first term accounts for previously infected agents who have not recovered, and the second term accounts for new infections. The validity of the former Markovian equations is shown in the Supplementary Material by comparing their results with those obtained from Monte Carlo simulations.\\

{\em Epidemic Threshold.-} One of the most important quantities in the study of epidemic spreading is the epidemic threshold $\lambda_c$, i.e., the minimum value of the infectivity $\lambda$ necessary for the epidemic to proliferate when a single infected agent is introduced in a population of fully susceptible agents. From its definition is clear its close connection to the basic reproductive ratio ${\mathcal R}_0$, which is the number of secondary cases that the former infected individual will cause in the population during their entire infectious period. Our goal now is to analyze how these metrics are influenced by both the mobility $p$ and return probability $r$, thus providing a better understanding of the differences between exploratory and commuting behaviors and their role for the proliferation or extinction of diseases.

Starting from Eq.~(\ref{eq:probab}), we linearize the probability of contagion as $P_i^D(t) \approx \lambda\sum_l I_{li}^D(t)$. Plugging the latter expression into Eq.~(\ref{eq:markov1}) and keeping linear terms, one can rewrite the time evolution of infected individuals as: 
\begin{equation}
    \mathbf{I}(t+1) \approx (1 - \mu) \mathbbm{N}\mathbbm{D}\mathbf{I}(t)\ + \lambda \mathbbm{N} \mathbbm{C} \mathbbm{D} \mathbf{I}(t)\ ,
\label{eq:linear}
\end{equation}
where we introduce $\mathbbm{C}$, the \textit{contact} operator, defined mathematically as:
\begin{equation}
\left(\mathbbm{C}\mathbf{I}\right)_{ij} = n_{ij}^D \sum_l I_{li}^D(t)\;.
\end{equation}

The basic reproduction number can be readily calculated by means of the \textit{next-generation} operator~\cite{diekmann1990definition,diekmann2010construction}. To this aim, we reorganize the terms of Eq.~(\ref{eq:linear}) to distinguish those corresponding to the creation of new cases $\mathbbm{T}$ from those involving transitions between the infectious states $\mathbbm{V}$: 
\begin{equation}
    \mathbf{\dot{I}} = \left( \underbrace{(1 - \mu)\mathbbm{N}\mathbbm{D} - \mathbbm{1}}_{\mathbbm{V}} + \underbrace{\lambda \mathbbm{N} \mathbbm{C} \mathbbm{D}}_{\mathbbm{T}} \right) \mathbf{I}(t)\;,
\end{equation}
where we have defined $\mathbf{\dot{I}}= \mathbf{I}(t+1)-\mathbf{I}(t)$. The basic reproduction ${\mathcal R}_0$ is the spectral radius, $\rho$, of the next-generation matrix $\mathbbm{K} = -\mathbbm{T}\mathbbm{V}^{-1}$. Following this result, we can derive $\lambda_c$ as the critical infectivity for which ${\mathcal R}_0 = 1$, resulting in
\begin{equation}
    \lambda_c = \dfrac{1}{\rho\left(\left(\mathbbm{N} \mathbbm{C} \mathbbm{D}\right)\left( \mathbbm{1} - (1 - \mu)\mathbbm{N}\mathbbm{D} \right)^{-1}\right)}\;.
    \label{eq:threshold}
\end{equation}
This solution provides the epidemic threshold as a function of the parameters $p$, $r$ and $\mu$. In the Supplementary Material we provide the explicit expressions of operators $\mathbbm{N}$, $\mathbbm{C}$ and $\mathbbm{D}$ that enable the calculation of $\lambda_c$ as well as the validation of Eq.~(\ref{eq:threshold}) in different scenarios.

\begin{figure}[t!]
\includegraphics[width=0.85\linewidth]{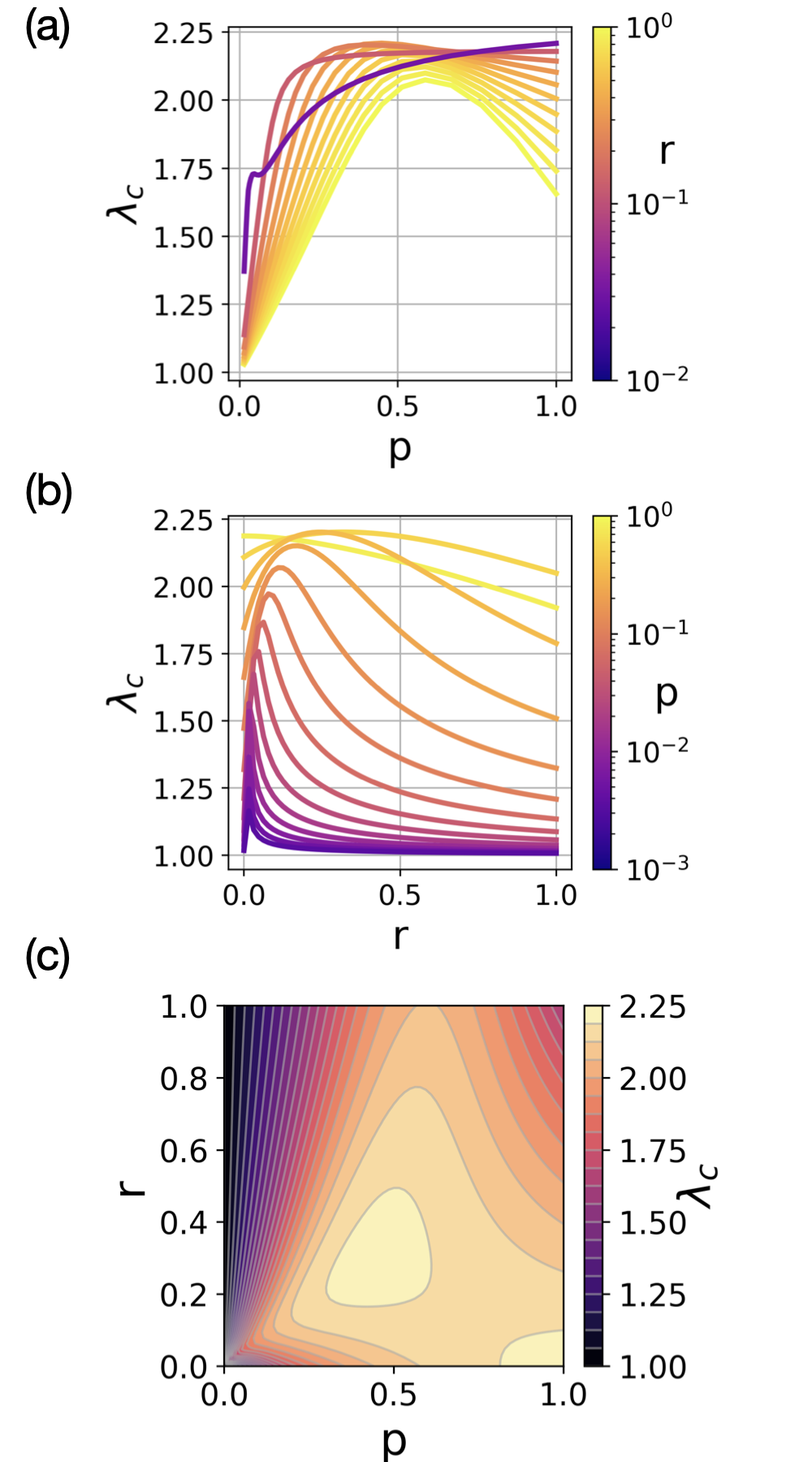}
\caption{\label{Fig2}(a) Epidemic threshold as a function of mobility $p$ for different values of the return probability $r$ in a BA network consisting of $50$ nodes and $\langle k\rangle =8$. (b) Epidemic threshold as a function of $r$ for different values of the probability $p$ in the same network as (a). The combination of both panels is shown in (c). In all the panels the threshold $\lambda_c$ is normalized with respect to $\lambda_0$, defined as the epidemic threshold for the static state ($p = 0$ and $r = 1$). 
}
\end{figure}

\begin{figure}[t]
\includegraphics[width=0.95\linewidth]{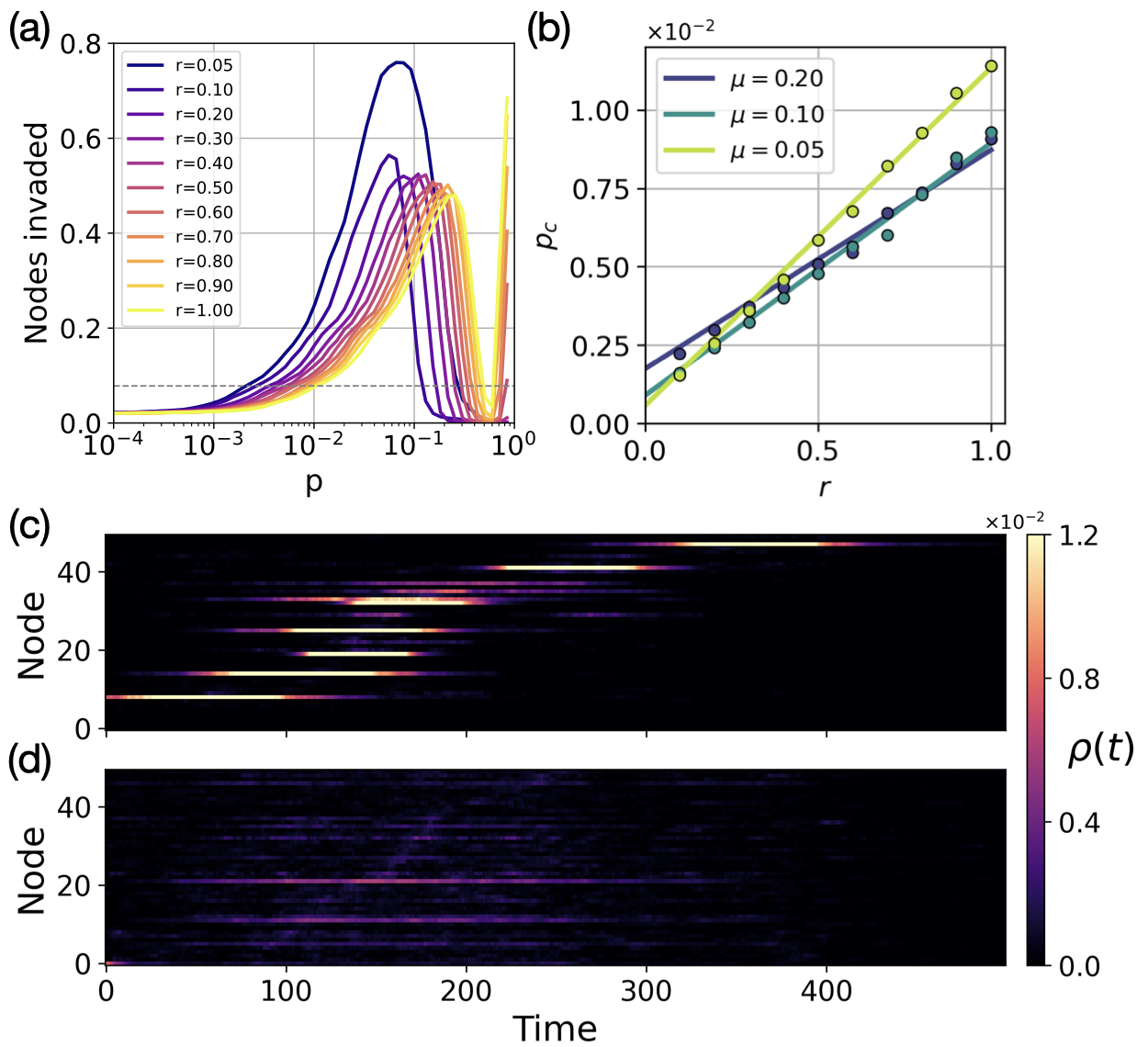}
\caption{\label{Fig3}(a) Proportion of nodes of the BA network that have been affected by the disease (reaching an attack rate of at least 2.5\%) as a function of $p$ and different values of $r$. The results correspond to an average across 500 Monte Carlo simulations with $\lambda= 2 \lambda_c(p=0)$ and $\mu = 0.2$. The dashed line represents the value of $f^{\star}=0.08$ chosen to calculate the threshold $p_c$.  (b) Invasion threshold $p_c$ as a function of $r$ and for different values of $\mu$. The linear regression has been fitted following equation \ref{eq:invasion} for each value of $\mu$. (c-d) Time evolution of infections across patches for $p = 0.01$ (c) and $p = 0.2$ (d) when $r=0.5$.}
\end{figure}

To evaluate the impact of mobility and recurrence on disease propagation, we use a Barab\'asi–Albert (BA) network with \( N=50 \) patches and average degree \(\langle k\rangle=8\). The initial population distribution \( n_{ij}(0) \) is randomly assigned, while edge weights \( W_{ij} \) are also randomly assigned considering that $\sum_{j=1}^{N}W_{ij}=1 $. Figure~\ref{Fig3}.a shows the epidemic threshold \( \lambda_c \) as a function of \( p \) for various \( r \) values. The \textit{epidemic detriment} effect \cite{gomez2018critical} is observed across all \( r \) values, where increased mobility (\( p>0 \)) initially raises the threshold. Notably, the nontrivial relationship \( \lambda_c(p) \) changes with \( r \); for large \( p \), \( \lambda_c(p) \) may increase or decrease depending on whether the flows are more exploratory or recurrent. Consequently, fixing \( p \) yields distinct \( \lambda_c(r) \) curves, as seen in Fig.~\ref{Fig2}.b. From these plots, we observe that promoting commuting results in a lower epidemic threshold, being this decline with $r$ monotonic primarily at high mobility levels.

The combined behaviors in Fig.~\ref{Fig2}.a-b are presented in Fig.~\ref{Fig2}.c as the surface \( \lambda_c(p,r) \). The contour plot reveals that the largest value of $\lambda_c$ is that of the pure random walk limit ($p=1$ and $r=0$) while in the static scenario ($p=0$) the minimum value of $\lambda_c$ takes place. Thus, from this region of low $p$ and high $r$ any increase of the mobility and/or the hops before coming back to the residence may suppress disease spread. In fact, by increasing further $r$ and $p$ one reaches a local maximum close to the center of the parametric plane, 
pinpointing that it is a trade off between commuting and exploratory patterns what confers more epidemic robustness to the system.\\

{\em Invasion threshold.}- So far, we have examined how mobility and commuting influence the critical properties of the transition from disease-free to epidemic states. 
However, in real epidemic scenarios, we often face a supercritical regime in which the high infectiousness of the circulating pathogen makes all the patches prone to develop local outbreaks. In such scenario, the role of mobility for the global epidemic outbreak becomes more nuanced. In particular, low (high) inter-region traffic in a metapopulation can inhibit (promote) the subsequent spread of local outbreaks among subpopulations, ultimately driving it toward an extinction (pandemic) scenario.

To capture this drift from local to global epidemic states one can measure the {\em invasion threshold}~\cite{colizza2007invasion}, defined as the lowest mobility value $p_c$ necessary for a local outbreak to invade a population in a supercritical regime. As this threshold comes from the stochastic nature of contagion/mobility processes, we must perform Monte Carlo simulations to obtain its value. In these simulations, we set $\lambda= 2 \lambda_c(p=0)$, $\mu=0.10$ and generate 500 outbreaks, each starting from a randomly selected patch. Figure~\ref{Fig3}.a shows the average fraction of invaded patches in an outbreak as a function of both $r$ and $p$. Focusing on the region $p\ll1$, we observe more widespread diseases as we increase $p$ and decrease the return probability $r$. This occurs because, in this regime, mobility barely alters the epidemic threshold (see Fig.~\ref{Fig2}.c) while fostering the dissemination of the disease across the system.

Quantitatively, we identify $p_c$ as the value for which the fraction of invaded patches surpasses a given threshold $f^{\star}$ (dashed line in Fig.~\ref{Fig3}.a). Figure~\ref{Fig3}.b represents the dependence of $p_c$ the return probability $r$ for different recovery probabilities $\mu$. We show that the invasion threshold always increases with $r$, being the effect of the latter more pronounced for long infectious periods (small $\mu$ values). An estimation of the dependence of $p_c$ with $r$ can be made by considering  
that the number of new cases seeded by a traveling contagious individual is proportional to the time the agent can navigate randomly the network without either recovering or coming to their original node (in which the initial local outbreak takes place). With this assumption and after some calculations (see SM), the threshold $p_c$ should satisfy:
\begin{equation}\label{eq:invasion}
    p_c  \sim (\mu^{-1} - 1)r\;,
\end{equation}
which captures the linear scaling with $r$ shown in Fig.~\ref{Fig3}.b.

Beyond $p_c$, the number of invaded patches  grows with $p$. However, this normal behavior only holds for $p\ll 1$, failing when the epidemic detriment phenomenon enters into scene. In particular, the increase of the epidemic threshold with $p$ (see Fig.\ref{Fig2}.c) reverts the growing tendency and, as reported in Fig.~\ref{Fig3}.a, for a large enough value of $p$ (that again depends on $r$) the global spread of the disease from local outbreaks decays and it is eventually suppressed. To further characterize this anomalous behavior, we represent two spatio-temporal epidemic trajectories generated by a pathogen considering, for the same value $r=0.5$, the intuitive $p=0.01$ (Fig.~\ref{Fig3}.c) and the anomalous $p=0.2$ (Fig.~\ref{Fig3}.d) regimes. Although both scenarios yield a similar fraction of invaded patches, we observe striking differences in the two trajectories. While for $p=0.01$ the invasion is described by a sequence of major local outbreaks, for $p=0.2$ outbreaks are minor and quasi-synchronous.\\

{\em Conclusions.-}  Similarly to social and biological systems~\cite{nowak_evolutionary_2006,miritello2013}, the combination of different strategies and behaviors has important consequences for the dynamical evolution of epidemics. In our case, we reveal that balancing commuting and exploratory mobility patterns can significantly influence disease propagation, achieving an optimal trade-off that maximizes the epidemic threshold and confers robustness at a global level. More importantly, linking global and local properties of disease spread, we find that high recurrence enhances local outbreak persistence in high-mobility regimes but may limit global outbreaks under low mobility. This result suggests that strategic and targeted travel restrictions, such as calibrating human flows 
to enhance commuting while limiting non-essential trips, are useful interventions aimed at containing local outbreaks before going global without inducing high socioeconomic costs~\cite{haug2020}.\\

{\em Acknowledgments.-} P.V. and J.G.G. acknowledge support from Gobierno de Arag\'on y Fondo Social Europeo (grant E36-23R) and Ministerio de Ciencia e Innovaci\'on (grants PID2020-113582GB-I00 and PID2023-147734NB-I00). D.S.P. acknowledges support from Ministerio de Ciencia e Innovaci\'on through grants JDC2022-048339-I and PID2021-128005NB-C21.

\bibliography{main}

\newpage

\renewcommand{\figurename}{Supplementary Figure}
\renewcommand{\thefigure}{\arabic{figure}}
\renewcommand{\tablename}{Supplementary Table}
\renewcommand{\thetable}{\arabic{table}}
\setcounter{figure}{0}

\newpage
\
\newpage

\section*{Supplementary Material}
\subsection*{Appendix A: Validation through agent-based simulations}

The Markovian equations described in the main text allow for an analytical study and fast iteration over time of the system's state. However, they need to be validated with agent-based Monte Carlo simulations. These simulations account for all the possible microscopic processes governing the change of the epidemic status or the spatial diffusion of each agent. For each time step, the following processes are thus simulated:
\begin{itemize}
    \item First, during stage M, each agent randomly decides with probability $p$ to move from their current location $j$. If so, they choose their destination among the neighboring patches of $j$, according to the entries of the mobility matrix ${\bf W}$.
    \item Secondly, during the I stage, each susceptible makes contacts with all the infected individuals in their current location, with a probability $\lambda$ of becoming infected in each contact. Conversely, every infected agent has a random chance of becoming recovered with probability $\mu$.
    \item Lastly, during the R stage, every agent returns to their original patch with probability $r$.
\end{itemize}

The simulations run until reaching the steady state of the SIR dynamics. Such state is characterized by the absence of infected individuals once the epidemic wave have passed. To quantify the extent of the epidemic one measures the occupation of the recovered compartment in the steady state, $R(\infty)$, also referred to as the attack rate of the outbreak.

Supplementary Figure 1 represents how the average attack rate, obtained considering $50$ Monte Carlo simulations, depends on the infectivity of the pathogen $\lambda$ for different $(p,r)$ values governing the flows of individuals across the metapopulation. Supplementary Figure 1 also illustrates the fair agreement between these results and those obtained by the numerical integration of the Markovian equations, thus validating our framework.
\begin{figure*}[h!]
\centering
\includegraphics[width=0.78\linewidth]{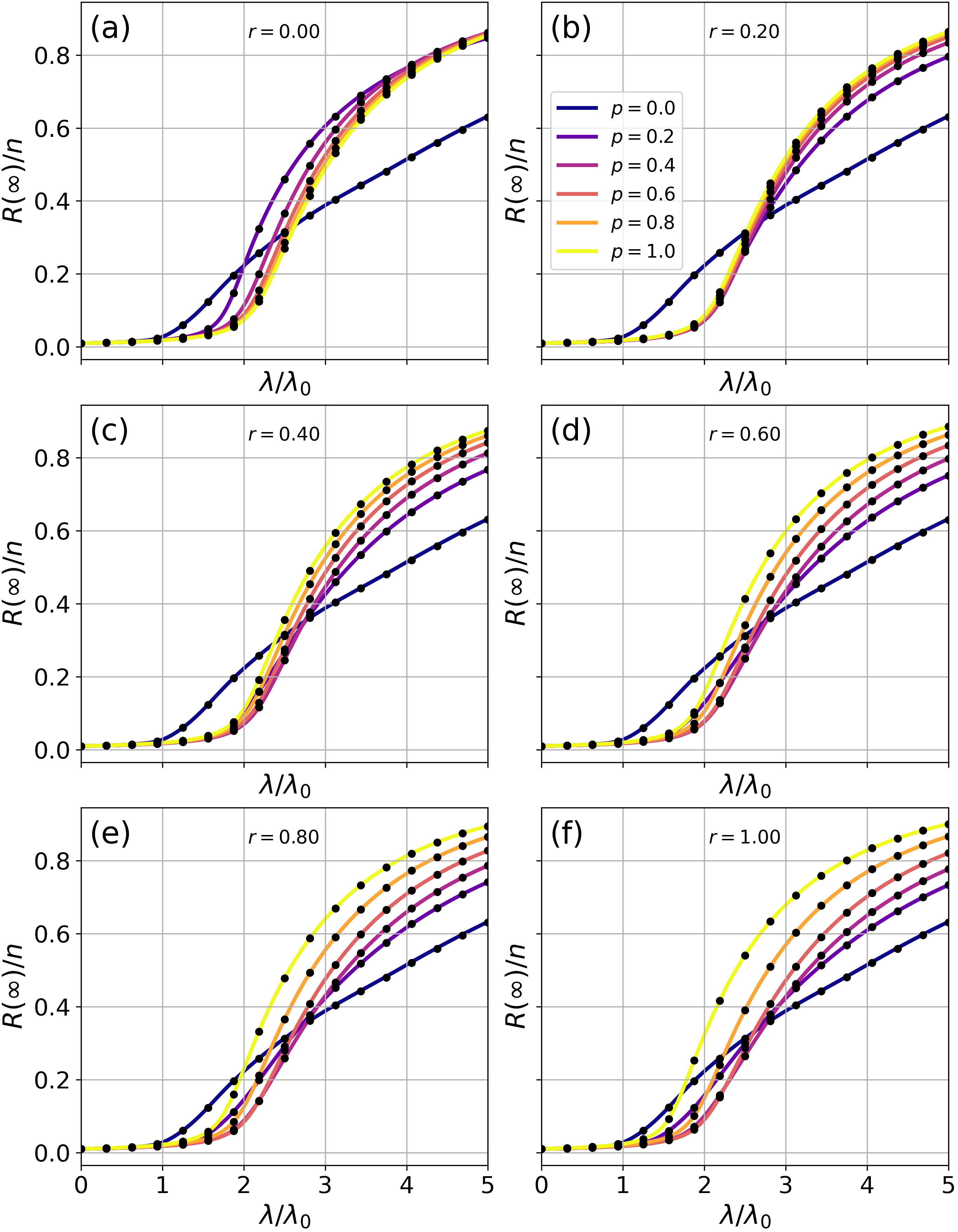}
\caption{\label{SuppFig1} (a-f) Validation of the Markovian equations in a Barab\'asi–Albert (BA) network, consisting of $50$ patches, an average degree of \(\langle k\rangle=8\) and a total population of $n=368000$ individuals. Each panel represents the attack rate of the outbreak $R(\infty)/n$ as a function of the infectivity $\lambda$ on the horizontal axis, normalized to the epidemic threshold in the absence of mobility as $\lambda_0 = \lambda_c(p=0)$. The recovery rate is $\mu = 0.20$, the mobility $p$ is represented by the color code (see legend in panel b) and the return rate $r$ is (a) $0$, (b) $0.20$, (c) $0.40$, (d) $0.60$, (e) $0.80$ and (f) $1$. The lines represent the results of iterating the Markovian equations and the points show the average number of recovered agents across $50$ simulations each.}

\end{figure*}

\subsection*{Appendix B: Movement, return and contact operators}

As shown in the main text, the linear operators operate over a vector of components $\mathrm{v}_{ij}$ as follows:
\begin{align*}
    &\left(\mathbbm{D} \mathbf{v}\right)_{ij} = (1 - p) \mathrm{v}_{ij} + p \sum_k W_{kj}\mathrm{v}_{ik}\;,\\
    &\left(\mathbbm{C}\mathbf{v}\right)_{ij} = n_{ij}^D \sum_k \mathrm{v}_{ki}\;,\\
    &\left(\mathbbm{N} \mathbf{v}\right)_{ij} = \delta_{ij}\sum_k r \mathrm{v}_{ik} + (1 - r) \mathrm{v}_{ij}\;.
\end{align*}
Meaning their explicit forms are:
\begin{align*}
    &D_{ij,kl} = \delta_{ik} \left( (1 - p) \delta_{lj} + p W_{lj} \right)\;,\\
    &C_{ij,kl} = \delta_{jl} n_{ij}^D\;,\\
    &N_{ij,kl} = \delta_{ik} \left( \delta_{lj} (1 - r) + r\delta_{ij}  \right)\;.
\end{align*}
where $\delta$ is the Kronecker delta.

\subsection*{Appendix C: Numerical Validation of the analytical expression of the epidemic threshold}
We validate the expression of the epidemic threshold obtained by computing it for different values of $r$ and $p$, and plotting it against the result of iterating the deterministic equations, which have already been validated using agent-based simulations, for a range of infectivities $\lambda$. Supplementary Figure 2 shows how the match between the separation of states and the epidemic threshold is exact.
\begin{figure*}[h!]
\centering
\includegraphics[width=0.80\linewidth]{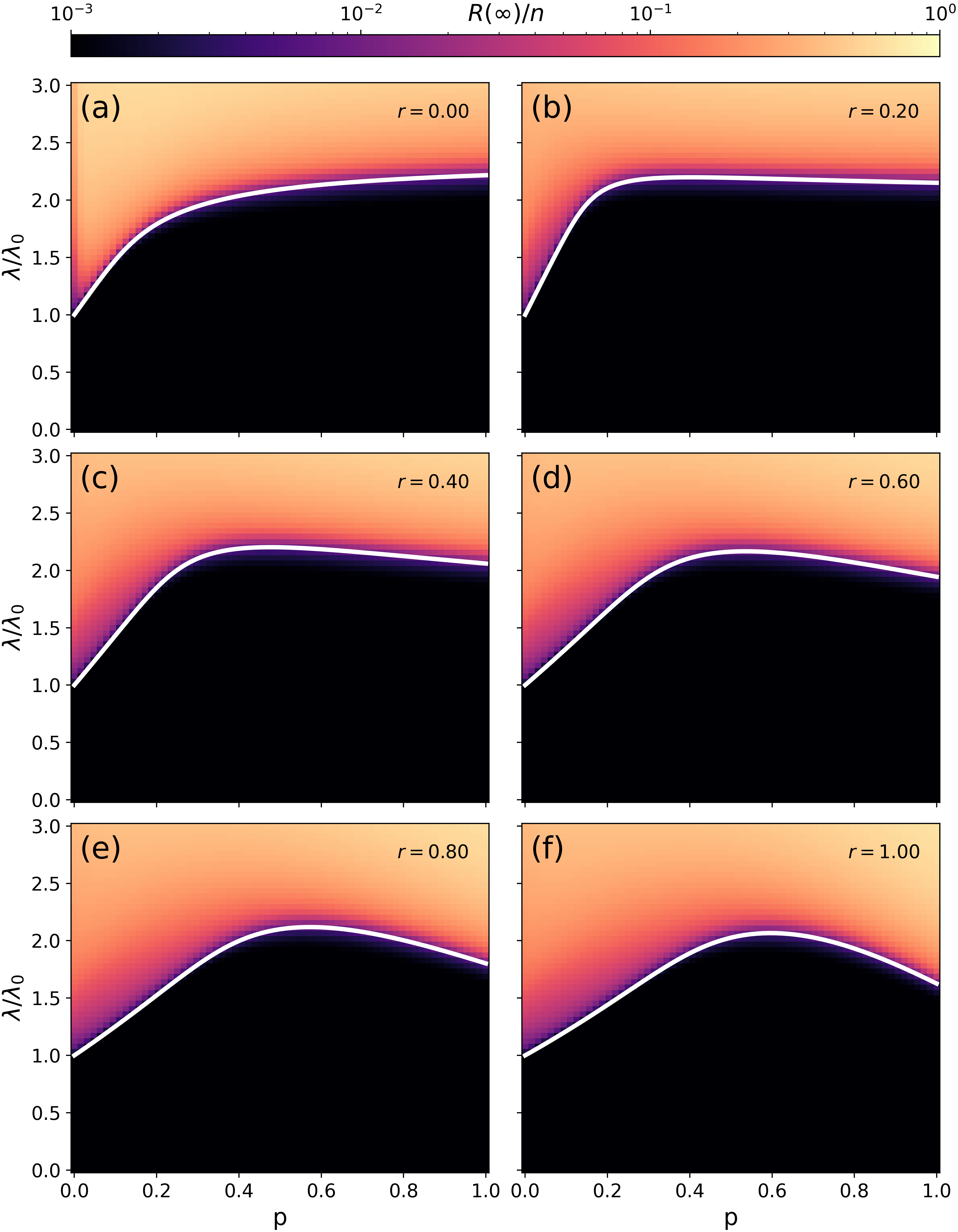}
\caption{\label{SuppFig2}(a-f) Epidemic threshold computed for different values of $p$ and $r$ (white continuous lines) compared to the attack rate as a fraction of the population obtained by iterating the deterministic equations (color code).}
\end{figure*}
\ 

\subsection*{Appendix D: Invasion threshold}

The Monte Carlo simulations show a clear relationship between the invasion threshold and the return rate $r$, and a first-order expression can be found after making some assumptions. The probability of a node  experiencing a local outbreak from a seed of infected agents is 
\begin{equation}\label{eq:P_inv}
\pi_{outbreak} = 1 - (1/R_0)^s\;,
\end{equation}
where $s$ is the number of local agents that have been infected by contagious individuals that come from another patch. This number is calculated as the number of secondary cases that are created by the roaming agents $\alpha$, or
\begin{equation}
    s = \alpha R_0'.
\end{equation}
$R'_0$ is different from the local $R_0$ because the time these agents spend at the patch being invaded is lower, as there is a chance that they return to their residence too soon. This means that the time they spend infecting others is $\mu + r(1 - \mu)^{-1}$ instead of $\mu$. As a result, the effective number of secondary cases they create is
\begin{equation}
    R'_0 = R_0 \frac{\mu}{\mu + r(1 - \mu)}
\end{equation}

Going back to equation \ref{eq:P_inv}, and assuming $R_0$ close to one, which is the case for many patches in the metapopulation, a first order approximation for the probability of outbreak can be made with $R_0 = 1 + \varepsilon$:
\begin{multline}
    \pi_{outbreak} = 1 - \left(\frac{1}{1+\epsilon}\right)^s\approx 1-(1-\epsilon)^s\approx \\
    1-(1-\varepsilon s) = \varepsilon s=(R_0 - 1)s\;.
\end{multline}
Considering now the expression for the number of infected seeds that got the disease from agents belonging to other patches:
\begin{equation}
    \pi_{outbreak} \approx (R_0 - 1)\frac{R_0\alpha\mu}{\mu + r(1 - \mu)}\;,
\end{equation}
where $\alpha$ is proportional to the mobility $p$, since it accounts for the number of contagious agents coming to the node from another one where there has been an outbreak. Therefore, the invasion threshold should obey $p_c\sim 1 + r(\mu^{-1} - 1)$.

\end{document}